\documentclass[aps,pra,a4paper,superscriptaddress,twocolumn,nofootinbib]{revtex4-1}

\usepackage{graphicx}
\usepackage[utf8]{inputenc}
\usepackage{float}

\usepackage{geometry}
\usepackage{amsmath}

\usepackage{xcolor}
 \definecolor{FAUgreen}{rgb}{0, .59, .46}
 \definecolor{FAUblue}{rgb}{0, .2, .4}

\usepackage{hyperref}
\hypersetup{
  pdfauthor = {B. Srivathsan, M. Sondermann},
  pdftitle = {Measuring the temperature and heating rate of a single ion by imaging},
  allbordercolors = {FAUgreen},
  urlbordercolor = {FAUblue}
}

\begin{document}

\title{Measuring the temperature and heating rate of a single ion by imaging}

\author{Bharath Srivathsan}
\affiliation{Friedrich-Alexander-Universit\"at Erlangen-N\"urnberg (FAU),
  Department of Physics, Staudtstr. 7/B2, 91058 Erlangen, Germany}
\affiliation{Max Planck Institute for the Science of Light,
Staudtstr. 2, 91058 Erlangen, Germany}
\affiliation{Centre for Cold Matter, Blackett Laboratory, Imperial College London,
  Prince Consort Road, London SW7 2AZ, United Kingdom}
\author{Martin Fischer}
\affiliation{Friedrich-Alexander-Universit\"at Erlangen-N\"urnberg (FAU),
  Department of Physics, Staudtstr. 7/B2, 91058 Erlangen, Germany}
\affiliation{Max Planck Institute for the Science of Light,
Staudtstr. 2, 91058 Erlangen, Germany}
\author{Lucas Alber}
\affiliation{Friedrich-Alexander-Universit\"at Erlangen-N\"urnberg (FAU),
  Department of Physics, Staudtstr. 7/B2, 91058 Erlangen, Germany}
\affiliation{Max Planck Institute for the Science of Light,
Staudtstr. 2, 91058 Erlangen, Germany}
\author{Markus Weber}
\affiliation{Friedrich-Alexander-Universit\"at Erlangen-N\"urnberg (FAU),
  Department of Physics, Staudtstr. 7/B2, 91058 Erlangen, Germany}
\affiliation{Max Planck Institute for the Science of Light,
Staudtstr. 2, 91058 Erlangen, Germany}
\author{Markus Sondermann}
\email{markus.sondermann@fau.de}
\affiliation{Friedrich-Alexander-Universit\"at Erlangen-N\"urnberg (FAU),
  Department of Physics, Staudtstr. 7/B2, 91058 Erlangen, Germany}
\affiliation{Max Planck Institute for the Science of Light,
Staudtstr. 2, 91058 Erlangen, Germany}
\author{Gerd Leuchs}
\affiliation{Friedrich-Alexander-Universit\"at Erlangen-N\"urnberg (FAU),
  Department of Physics, Staudtstr. 7/B2, 91058 Erlangen, Germany}
\affiliation{Max Planck Institute for the Science of Light,
Staudtstr. 2, 91058 Erlangen, Germany}
\affiliation{Department of Physics, University of Ottawa, Ottawa,
  Ont. K1N 6N5, Canada}

\begin{abstract}
We present a technique based on high resolution imaging to measure the
absolute temperature and the heating rate of a single ion trapped at
the focus of a deep parabolic mirror. We collect the fluorescence
light scattered by the ion during laser cooling and image it onto a
camera. 
Accounting for the size of the point-spread function and the
magnification of the imaging system, we determine the spatial extent
of the ion, from which we infer the mean phonon occupation number in
the trap. 
Repeating such measurements and varying the power or the
detuning of the cooling laser, we determine the anomalous heating rate.
In contrast to other established schemes for measuring the heating
rate, one does not have to switch off the cooling but the ion is
always maintained in a state of thermal equilibrium at temperatures
close to the Doppler limit.  
\end{abstract}
\maketitle

\section{Introduction}
In many atomic physics and quantum optics experiments, the temperature
of the atoms under investigation plays a critical
role~\cite{Gibble1993, Ludlow2015, Cronin2009, Bloch2008, HAFFNER2008,
  Saffman2010}.  
From fundamental tests to quantum information applications, cooling the atoms to ultra-low temperatures has become a prerequisite. 
To this end, laser cooling and trapping of atoms has become an indispensable tool in many labs. 
Furthermore, measuring the temperature of the cold atoms becomes important, for instance
to understand the physics of the cooling mechanism, or to disclose
additional sources of heating and thermal decoherence in the experiment. 
Depending on the type of the trapping and cooling method employed, several thermometry techniques have been developed. 

In the case of trapped ions, the most common way to determine the
temperature is to measure the sideband absorption
spectrum~\cite{Diedrich1989, Roos1999}. This technique requires the
ion to be cooled close to the motional ground state of the trap, and
therefore is used in combination with ground state cooling schemes such
as Raman side-band cooling~\cite{Monroe1995} or cooling employing electromagnetically induced
transparency~\cite{Roos2000}.  
Outside the resolved sideband regime, various techniques exist.
One way is to measure the Doppler broadening of the atomic transition due to
the motion of the ion \cite{Wineland1979, Wineland1981}. The accuracy
of this approach relies on the ability to distinguish the
Lorentzian spectrum of an atomic transition from
the Gaussian spectrum. 
The Doppler broadening at sub-mK temperatures is small compared to 
the natural linewidth of the typically used transitions. 
Therefore, the statistical uncertainties in data evaluation
prevent an accurate determination of the temperature
close to the Doppler limit. 
Particularly, in the Lamb-Dicke Regime the first order Doppler effect 
is suppressed, and only the higher order Doppler shifts that are much 
weaker can be observed~\cite{Dicke1953, Itano1988}.

In order to enable fast and accurate determination of the temperature of a trapped ion 
at mK temperatures, thermometry by imaging the spatial extent of an ion 
has been demonstrated~\cite{Norton2011, Knunz2012}. The accuracy of
this method was limited only by the imaging resolution and the images' 
signal-to-noise ratio.  In this article, we present the thermometry measurement 
of a single ion by imaging via a deep parabolic mirror. 
Our improved resolution and high collection efficiency~\cite{Maiwald2012} allows us to
determine the absolute temperature close to the Doppler limit 
more accurately in comparison to previous demonstrations of this
technique. Furthermore, it opens up the possibility to measure
temperatures below the Doppler limit, which has until now been
possible only by the resolved sideband method or by similarly involved
methods such as the one based on interference of fluorescence
photons~\cite{slodicka2012}.

In addition to the absolute temperature, the heating rate is  another 
important figure of merit in thermometry of trapped ions~\cite{Deslauriers2006, Brownnutt2015}. 
In the resolved sideband regime, sideband themometry is generally employed to measure 
the heating rate. Outside this regime, the heating rate is traditonally 
measured from the time-resolved fluorescence rate of the ion during 
the Doppler cooling process~\cite{Epstein2007}. 
A similar technique combining the imaging
approach and the time-resolved scattering method to determine the
heating rate was recently demonstrated~\cite{boldin2018}. Both these
techniques involve heating up the ion to temperatures at-least a few
orders of magnitude above the Doppler limit, and therefore depend on
several simplifying assumptions about the system. 
We present an alternative way to determine the heating rate of a
single ion employing the imaging approach while varying the
cooling laser power or its detuning. The advantage of our technique is
that for every measurement point the ion is maintained in a state of
thermal equilibriumduring the entire measurement sequence.

\section{Theory}

The temperature $T$ of an ion in an harmonic trap under weak confinement conditions
can be approximated as $T \approx \bar{n} \hbar \omega / k_B$, where $\bar{n}$ 
is the average excitation number of the harmonic oscillator, $\omega$ is the trap frequency and 
$k_B$ is the Boltzmann constant~\cite{Knunz2012}.
$\bar{n}$ in turn is related to the RMS spread $\sigma_i$ of the ion in position space as
\begin{equation}
  \label{eq:sigma}
  \sigma_{i}= \sqrt{\frac{\hbar}{2 m \omega} (2 \bar{n} + 1)} \quad,
\end{equation}
where $m$ is mass of the ion. In the experiment, we measure $\sigma_{i}$ by imaging the ion, 
and thereby determine $\bar{n}$ and $T$.

In order to determine the heating rate $\zeta$ induced by external
factors, we use a simple model of laser cooling~\cite{leibfried2003} 
which neglects additional heating or cooling 
due to micromotion. Since micromotion is well compensated in 
our experiment, this model is well suited to describe the cooling process. 
Cooling as well as heating induced by interaction of the ion with the
cooling light is governed by scattering of photons from the
near-resonant cooling beam.
The steady state scattering rate of these photons is given by
$\Gamma\cdot \rho_{ee}$, where $\Gamma$ is the spontaneous emission
rate of the cooling transition and 
$\rho_{ee}=\Omega^2/(4\Delta^2 + \Gamma^2 + 2\Omega^2)$ is the steady
state excitation of the ion.
$\Delta$ is the detuning of the laser from the atomic resonance and $\Omega$
is the Rabi frequency. 
Particles confined in harmonic traps can have anisotropic temperature, 
depending on the angle made by the cooling beam with the trap axes~\cite{Norton2011}. 
To include this effect in our model, we define an effective k-vector,
$|\vec{k}_{\text{eff}}|\,=\,|\vec{k}|\,\cos{\alpha}$, where $\alpha$ is the angle made by 
the cooling laser with a trap axis. 
The cooling rate along the chosen trap axis is given by
\begin{equation}
 \dot{E}_{c} = -\frac{8 \hbar|\vec{k}_{\text{eff}}|^{2} \Delta \Gamma}{\Omega^2} \frac{k_B T}{m} \rho_{ee}^{2} \quad
\end{equation}
The heating rate during the final stages of Doppler cooling can be approximated as
\begin{equation}
\dot{E}_{h} = \frac{\rho_{ee}\,\Gamma}{2\,m}[\hbar^2 |\vec{k}_\text{eff}|^2 + \xi \, \hbar^2 |\vec{k}|^2] + \zeta \quad
\end{equation}
The first term in the brackets corresponds to the momentum change along the trap axis due to absorption of a photon, 
while the second term corresponds to the momentum change due to spontaneous emission along this direction. $\xi$ is a geometry factor 
that originates from the spatial emission characteristics of the
scattered photons. In our experiment, we use a $\text{J}_\text{1/2}
\rightarrow \text{J}_\text{1/2}$  
transition with a nearly isotropic emission pattern. Therefore, we use a geometry factor of $\xi\,=\,\text{1/3}$. 
In addition, we use a constant factor $\zeta$ to include anamalous heating in the model.

The equilibrium temperature is reached when the heating and cooling rates are equal: 
$\dot{E}_{c}+\dot{E}_{h}=0$.
Below, we will measure $\sigma_i$ while either varying $\Omega$ or
$\Delta$. In both cases, we will obtain $\zeta$ by fitting our model to
the experimental data. 

\section{Experiment}
\label{sec:exp}

\begin{figure}[t]
\centering
\includegraphics[width=\columnwidth]{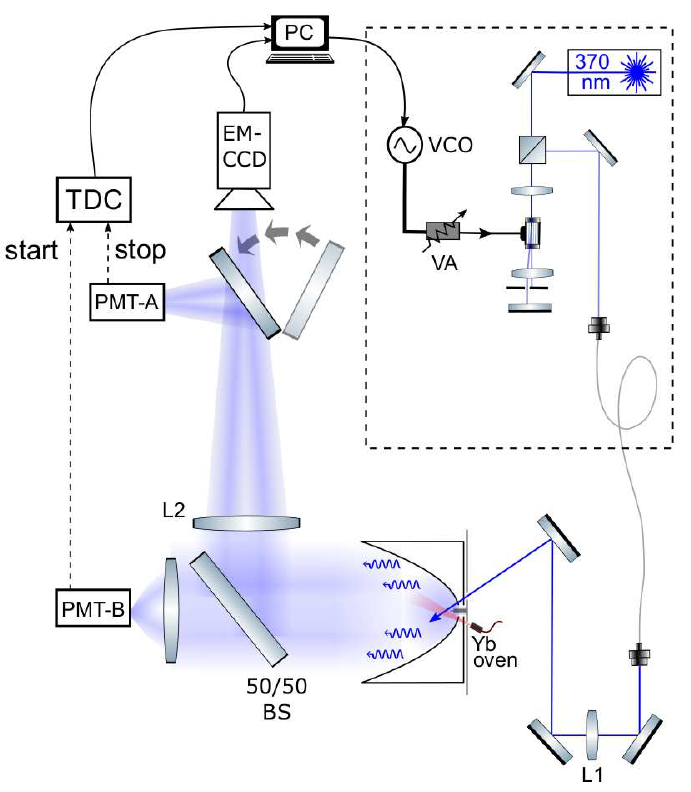}
\caption{ Schematic of our experimental setup. A stylus ion trap is
  used to trap a single $^{174}$Yb$^{+}$ at the focus of a deep
  parabolic mirror (PM).} 
\label{fig:exp}
\end{figure}

The schematic of our experimental setup is shown in
Figure~\ref{fig:exp}. We trap a single $^{174}$Yb$^+$ ion in the focal
region of a deep parabolic mirror using a stylus like ion
trap~\cite{Maiwald2009}.  
The trap is mounted on a XYZ piezo translation stage (PIHera P-622K058) 
with a positioning accuracy of about
$\pm$1\,nm. With the aid of the piezo stage, the ion can be positioned
and scanned in all three directions around the focal point of the
mirror. 
A 370\,nm frequency doubled diode laser (Toptica) is used for Doppler
cooling the ion. The detuning of the laser is tuned by using a
200\,MHz Accusto-Optic-Modulator (AOM), aligned in ``double-pass''
configuration, and driven by the amplified signal of a Voltage Controlled Oscillator
(VCO).  
The frequency shifted beam is coupled into a polarization maintaining
single mode optical fiber, and focused onto the ion using a 400\,mm
focal length lens (L1).  
The optical power of the cooling beam can be tuned by varying the RF
power supplied to the AOM using a Variable attenuator (VA). 
The parabolic mirror (focal length of 2.1\,mm) collimates the
fluorescence light scattered by the ion, and acts as an objective for
our imaging system.  
A 300\,mm focal-length lens (L2) along with a one-to-one telescope using 
lenses of focal length 50\,mm (not shown in Fig.~\ref{fig:exp}) is used to image the ion on an 
electron-multiplying charge-coupled device (EM-CCD) camera. 
A flip mirror (FM) directs the fluorescence photons instead to a
Photo-Multiplier-Tube (PMT-A).
The trap frequencies were measured to be 205\,kHz and 196\,kHz in the
lateral directions (X and Y), and 390\,kHz in the axial direction (Z) by applying 
AC fields to one of the trap electrodes. The cooling beam has an angle, $\alpha$ of 
71$^{\circ}$ with both the lateral trap axes.

\begin{figure}[t]
\centering
\includegraphics[width=\columnwidth]{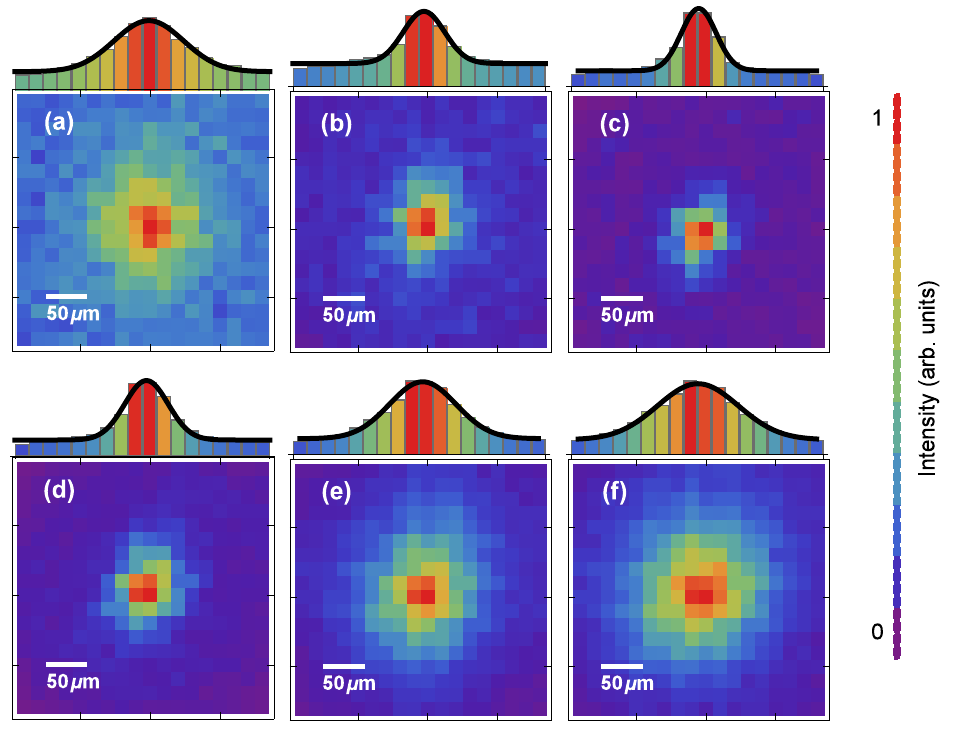}
\caption{\label{fig:images}
  Images of the single ion on the EM-CCD camera for Rabi frequencies
  $\Omega$ of  {\bf (a)} 0.012\,$\Gamma$, {\bf (b)} 0.025\,$\Gamma$,
  {\bf (c)} 0.23\,$\Gamma$, {\bf (d)} 1.1\,$\Gamma$, {\bf (e)}
  2.4\,$\Gamma$, and {\bf (f)} 3.3\,$\Gamma$. The pixel sum along the
  columns is shown as bars above each image. The solid (black) line
  shows the corresponding Gaussian fits used to determine the image
  width along this direction, $\sigma$.
  The color scale is normalized to the maximum intensity observed in
  each image, respectively.
  All images are recorded at a detuning $\Delta=-2\,\pi\times13\,\text{MHz}$.
}
\end{figure}

As introduced above, we measure the temperature of the ion by
determining the width of the image recorded on the EM-CCD camera. 
Due to the geometry of the trap electrodes
(cf. Ref.~\cite{Maiwald2012}),
the lateral trap axes make an angle of 45$^{\circ}$ with respect to
the edges of the pixel array of the camera. 
Therefore, we rotate the images by nearest neighbor interpolation to
make the the pixel array axis coincide with the trap  
axis~\footnote{ We use MATLAB {\em imrotate} function to rotate the images}.
For simplicity, we restrict the discussion to one spatial dimension.

First, we project the image onto the horizontal direction by summing over all
pixels in each column.
We then determine the RMS image width $\sigma$ from this projection by
using a 1D Gaussian fit.
The image recorded on the camera is a convolution of the imaging
point-spread function (PSF) and the ``true image''  of the
ion. Assuming both the PSF and the true image to be Gaussian spots,
the width of the recorded image can be approximated as  
\begin{equation}
\sigma = \sqrt{\sigma_\text{PSF}^2 + M^2 \sigma_{i}^2} \quad .
\end{equation}
In order to determine the temperature of the ion, we need to extract $\sigma_i$ from the measured $\sigma$. 
For this process the magnification $M$ of the imaging system as well
as the width $\sigma_\text{PSF}$ of the imaging PSF need to be known.
$M$ is measured by moving the ion in lateral directions, and
measuring the image shift on the camera as outlined in
App.~\ref{sec:magnify}, yielding $M=113\pm2$.
We use $\sigma_\text{PSF}$ as a free parameter in the 
fitting procedure discussed below. 

Several example images acquired at different Rabi frequencies $\Omega$
are shown in Fig.~\ref{fig:images}. The Rabi frequency is obtained
by one calibration measurement at fixed power (see App.~\ref{sec:rabi}). 
All other values of $\Omega$ are then calculated from the power of the
cooling laser and the power used in the calibration measurement.
As expected, the width of the images in
Fig.~\ref{fig:images} varies with the Rabi frequency $\Omega$.

\begin{figure}[t]
\centering
\vspace{-4mm}
\includegraphics[width=\columnwidth]{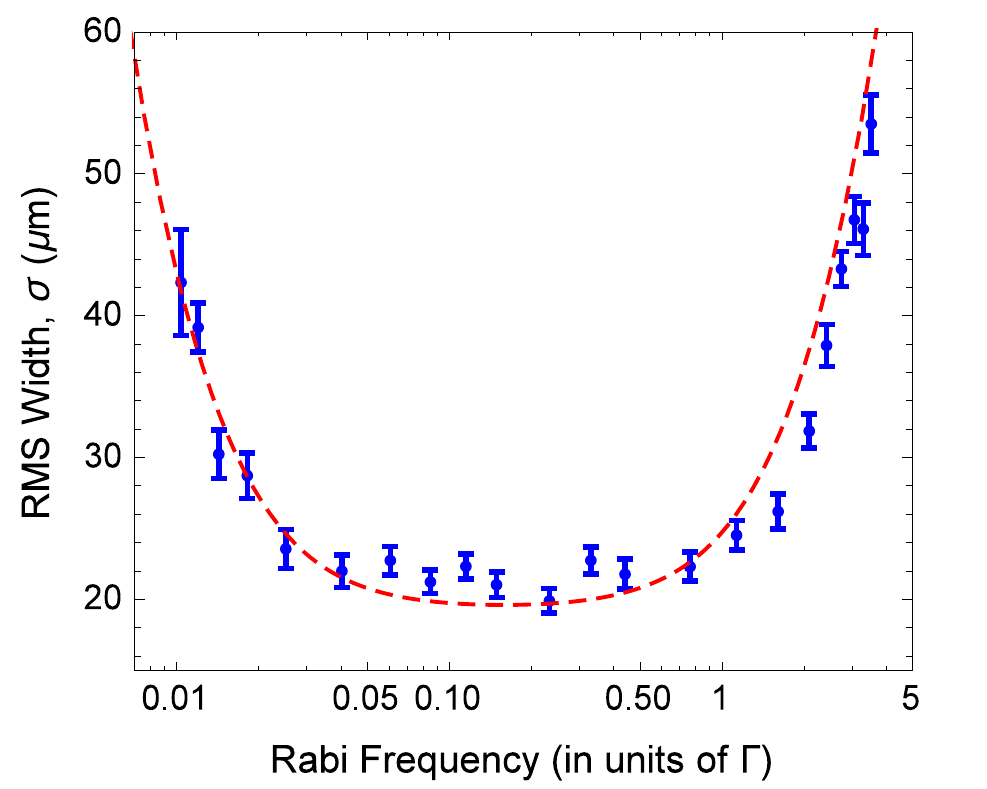}
\caption{Image RMS width along the horizontal direction as a function of
  Rabi frequency $\Omega$ at a fixed detuning of
  $\Delta/2\pi=-13\,\text{MHz}$.
  The dashed (red) represents a fit to the
  model described in the text. The only free parameter used in the fit is 
  the heating rate, $\zeta$. The predicted value of $\zeta$ is \,0.38$\pm$0.07\,quanta/ms.}
\label{fig:satplot}
\end{figure}

We now turn to the determination of the heating rate $\zeta$.
We fit our model to a set of image widths $\sigma$ by varying either $\Omega$
at fixed $\Delta$ or vice versa.
The free fit parameters are $\zeta$ and $\sigma_\text{PSF}$.
The result of the measurement and the fit for varying $\Omega$ at a
fixed detuning $\Delta/2\pi=-13\,\text{MHz}$ are
shown in Figure~\ref{fig:satplot}. 
The increase of the image width and thus the ion temperature at very
small Rabi frequencies indicates a non-zero excess heating.
The corresponding heating rate as determined from the fit is
0.38$\pm$0.07\,quanta/ms. The width of the imaging PSF is 
$\sigma_\text{PSF}$\,=\,6.6$\pm$2.7,$\mu$m, which is in good agreement with 
the expected PSF of $7.1\,\mu\text{m}$ determined from the simulations including the interferometrically measured 
aberrations of our parabolic mirror ~\cite{leuchs2008} (see App.~\ref{sec:PSF}).
The average phonon number $\bar{n}$ and thus the temperature of the
ion for any $\Omega$ can now be determined using these parameters.

\begin{figure}[t]
\centering
\includegraphics[width=\columnwidth]{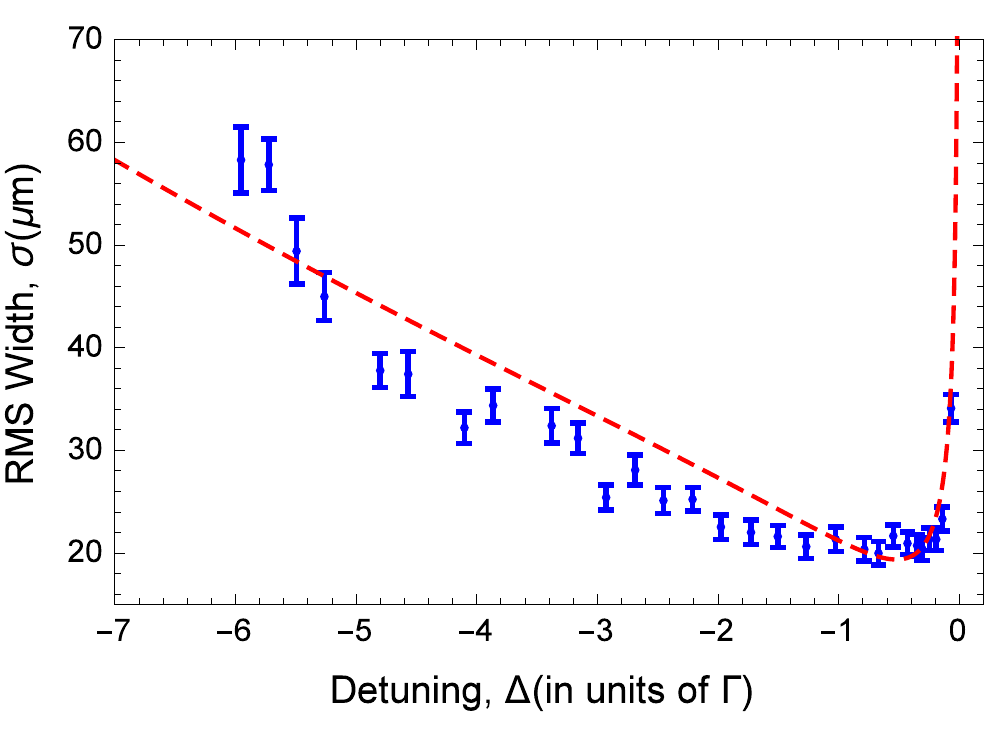}
\caption{Image RMS width along horizontal direction $\sigma$ as a function of
  detuning $\Delta$. The cooling beam power is fixed at a value corresponding 
  to the on-resonant Rabi frequency ($\Omega$) of 0.2\,$\Gamma$.
  The dashed (red) line represents a fit to the
  model. The heating rate $\zeta$ predicted by the fit is
  0.22$\pm$0.07\,quanta/ms.
  } 
\label{fig:detuningplot}
\end{figure}

The lowest measured $\sigma_i$ is 0.166$\pm$0.013\,$\mu$m, for
a Rabi frequency of $\Omega$\,=\,0.23\,$\Gamma$. This corresponds to a
mean phonon occupation number $\bar n$ of 97\,$\pm$\,15, and a
temperature of 950$\pm$147\,$\mu$K.
In the Doppler limit, the temperature according to
$k_\text{B}T \approx \hbar\Gamma/2$ is expected to be about
$T_\text{D}=\,470\,\mu\text{K}$.
Thus, the temperature of the ion is found to be about twice this Doppler limit, 
mainly due to a large angle between the cooling beam and the trap axis.

It can be seen that for the measured temperatures, the contribution of the 
PSF ($\sigma_\text{PSF}$) to the measured image width ($\sigma$) 
is much smaller than the contribution from spatial extent of the ion wavefunction ($\sigma_{i}$).
Hence, this method can be used for measuring even lower temperatures.
From the standard error estimates of $\sigma_\text{PSF}$ and $M$, 
we estimate the minimum measurable temperature with a 50\% relative error to be $\approx\,200\,\mu\text{K}$ 
\footnote{ This estimate is an absolute lower limit for our experimental parameters, 
assuming that the poissonian noise in image acquisition is negligible compared to uncertainities of PSF and magnification estimates.}, 
which is well below the standard Doppler limit.

An alternative way to measure the heating rate of the ion is to
measure $\sigma$ when varying the detuning $\Delta$.  
We fix the cooling beam power such that $\Omega$\,=\,0.2\,$\Gamma$. 
The detuning is varied by using the VCO, and the image width is measured as a function of the detuning. 
The result is shown in figure~\ref{fig:detuningplot}.
From a fit we extract a heating rate of
$0.22\pm0.07$\,quanta/ms, which is in fair
agreement with the previous measurement.

\section{Conclusion}
We have demonstrated a technique to measure the absolute temperature
of a single ion and its heating rate by measuring its spatial
probability distribution in the trap. The high resolution image of the ion
obtained by using our parabolic mirror as imaging tool allows us to
measure temperature close to the Doppler limit, indicating the potential to
perform thermometry below the Doppler limit. We have also measured the
heating rate in our trap, while the ion is constantly maintained in a
thermal equilibrium. Therefore, this technique might be useful for
traps exhibiting high anharmonicity or temperature dependent heating
rates. 

\section*{Acknowledgments}
G.L. acknowledges financial support from the European Research Council
(ERC) via the Advanced Grant 'PACART'.

\appendix

\section{Magnification Calibration}
\label{sec:magnify}

\begin{figure}
\centering
\includegraphics[width=\columnwidth]{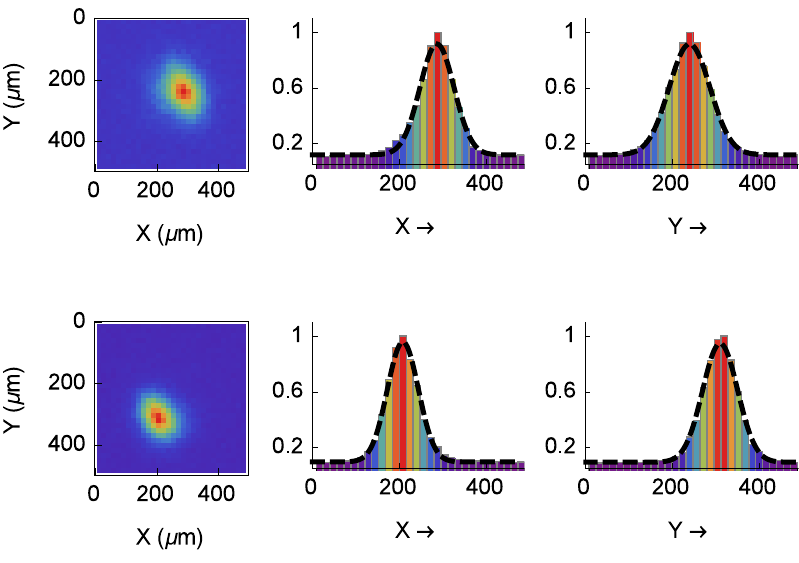}
\caption{{\bf(Left)} Images recorded on the EM-CCD camera for two positions
  of the ion, displaced in the by 635(2)\,nm in x-direction and 665(2)\,nm in y-direction. 
  {\bf(Middle)} Row scan of the images. Dashed line reperesents Gaussian fit function used to 
  determine the x centers. 
  {\bf(Right)} Column scan of the images and the corresponding fit functions used to determine 
  the y centers.
  The difference in the center position between the two images is 74.6$\pm$1.8\,$\mu$m and 
  72.5$\pm$1.8$\,\mu$m for x and y directions respectively.
}
\label{fig:magplot}
\end{figure}

To calibrate the magnification of the imaging system, we move the ion
in the lateral direction using the piezo stage, and record the images
on the camera as shown in figure~\ref{fig:magplot}. By comparing the
image shift to the object (ion) displacement, we determine the
magnification of our imaging system to be $M_x = 118(3)$ in the 
horizontal direction, and  $M_y = 109(3)$ in the vertical direction. 
The magnification along the trap axes is $M = \sqrt{(M_x^2+M_y^2)/2}=113(2)$.

\section{Imaging PSF}
\label{sec:PSF}

We simulate intensity distribution of a radially polarized doughnut mode 
focused by our parabolic mirror, include the interferometrically measured 
aberrations by a generalization of the method presented in \cite{Richards1959}.
We obtain a FWHM width of the intensity distribution of $148\,\text{nm}$ for our wavelength.
Including the magnification $M$ of our imaging system, this translates to an expected PSF $7.1\,\mu$m at 
the EMCCD camera.

\begin{figure}
\centering
\includegraphics[width=0.75\columnwidth]{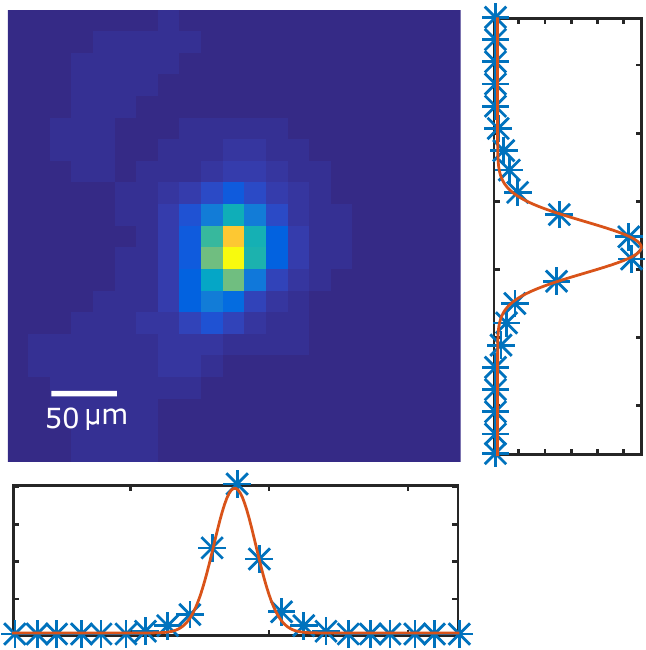}
\caption{Imaging PSF measured with a radially polarized doughnut
  beam. The plots show the row and column projections of the recorded image, and the solid red line is Gaussian fit.
The RMS width is determined from the fit to be
 12.3$\pm0.5$\,$\mu$m in the horizontal direction, and
 16.3$\pm0.5$\,$\mu$m in the vertical direction.} 
\label{fig:PSF_plot}
\end{figure}

Although also extractable from the fits presented in
Sec.~\ref{sec:exp}, we give an independent estimate of the size of the
PSF of our imaging system as a consistency check.
We generate a collimated radially polarized doughnut beam at the
wavelength of the $P_{1/2}\rightarrow S_{1/2}$ transition at 370\,nm
wavelength as described e.g. in Ref.~\cite{alber2017}.
This mode is focused by the parabolic mirror.
The rediverging beam is collimated again by the paraboloid and also
imaged onto the EM-CCD camera.
It thus passes the same optical elements as the fluorescence photons
detected during the temperature measurements.
The RMS width of the image on the EM-CCD chip is determined by 1D
Gaussian fits as shown in Figure~\ref{fig:PSF_plot}, yielding a
width of  $\sigma_\text{PSF}=12.3\pm0.5\,\mu\text{m}$ and 
$16.3\pm0.5\,\mu\text{m}$ for the horizontal and vertical
direction, respectively.

The $\sigma_\text{PSF}$ extracted from these measurements is also
influenced by residual aberrations stemming from the optical elements
used for preparing the doughnut beam.
Moreover, this beam is reflected twice at the surface of the
parabolic mirror. Thus, aberrations due to a non-perfect parabolic shape of the
mirror~\cite{alber2017,leuchs2008} are imprinted twice onto this beam.
The phase front of the fluorescence photons emitted by the ion
carries these aberrations only once.
Furthermore, the spatial mode of the laser used in that measurement is not
the same as the average spatial emission pattern of a $^{174}$Yb$^+$
ion emitting photons on the  $P_{1/2}\rightarrow S_{1/2}$ transition.
After collimation by the parabolic mirror, the intensity pattern of
the ion's fluorescence is of Lorentzian shape~\cite{Maiwald2009}. 
Therefore, the width of the PSF obtained in this measurement can be
considered as an upper bound for $\sigma_\text{PSF}$ in the
temperature measurements.

\section{Determination of the Rabi frequency}
\label{sec:rabi}

\begin{figure}[h]
\centering
\includegraphics[width=\columnwidth]{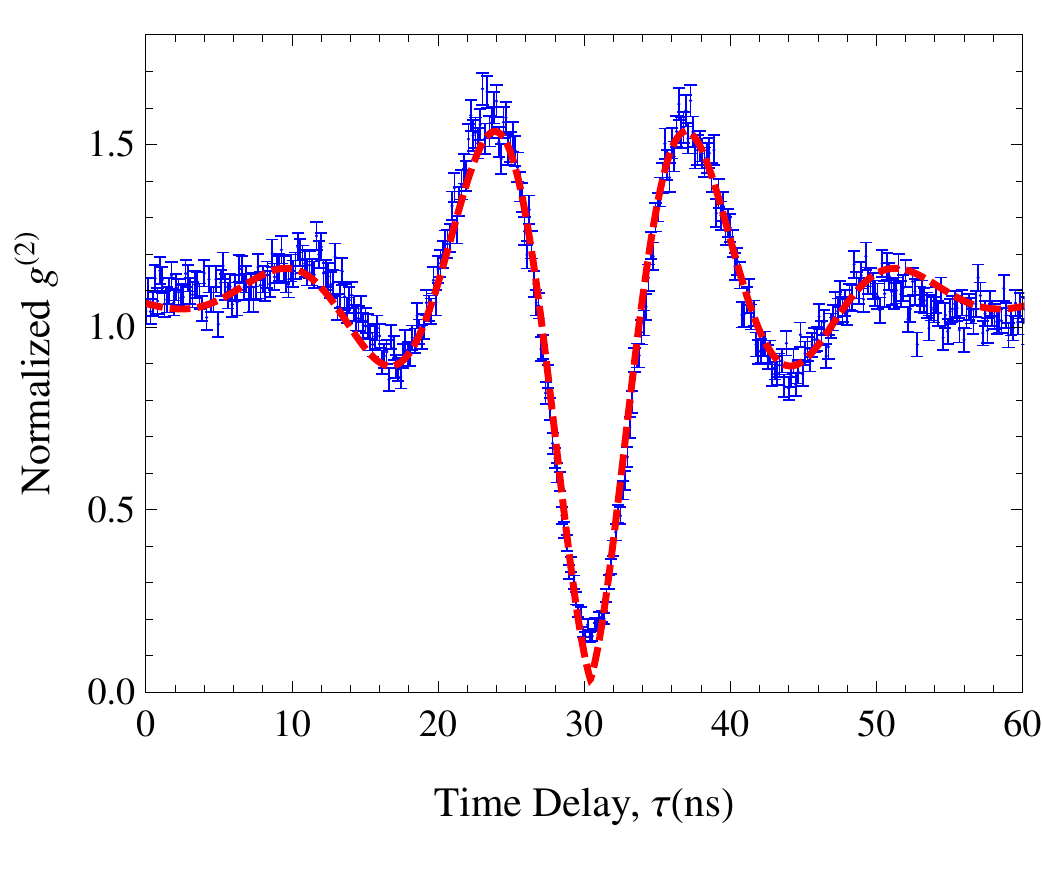}
\caption{\label{fig:rabi}
Normalized second order auto-correlation function of the ion
fluorescence for the cooling beam power of P\,=\,50\,$\mu$W. Delay 
in the minimum of $g^{(2)}$ is a result of an electronic delay between the 
the detection events registered by PMT-A and PMT-B.
The Rabi Frequency $\Omega'$ is determined from a fit of the measured
points to a function of the form $g^{(2)} (\tau) =
bg\,-\,A\,(\textrm{Cos}^{2}[\frac{\Omega}{2}(t-t_0)] -
\frac{1}{2})\,e^{\frac{-|t-t_0|}{\tau}} $. 
The resulting parameters
from the fit are: $bg$\,=\,1.08(0), $A$\,=\,2.09(3),
$\Omega'$\,=\,0.439(2)\,$\mu\text{s}^{-1}$, $t_0$\,=\,30.4(0)\,ns, and
$\tau$\,=\,8.16(16)\,ns. 
}
\end{figure}

In order to determine the temperature of the ion from the image size,
it is essential to precisely measure the on-resonance Rabi frequency
$\Omega$. We measure the effective Rabi frequency $\Omega'$ by
performing a Hanbury-Brown-Twiss type experiment on the fluorescence
photons.
The detuning of the cooling beam is fixed at $\Delta=-\Gamma/2$. 
The power of the cooling beam, as measured by a power meter (Ophir Nova II), is fixed at a
calibration value of P\,=\,50\,$\mu$W. 
This value is chosen such that $\Omega'\gg\Delta$, which makes it possible 
to observe Rabi oscillations in $g^{(2)}$ measurement within the decay time.
The on-resonance Rabi frequency is then determined from $\Omega'$
and $\Delta$, $\Omega=\sqrt{\Omega'^2 - (\Gamma/2)^2}=435(2)\,\mu\text{s}^{-1}$.

The fluorescent light from the
ion is split using a 50/50 beam splitter, and detected using two
Photo-Multiplier Tubes PMT-A and PMT-B. A Time-to-Digital Converter
(TDC) is used to measure a Start - Stop correlation histogram
between the PMT clicks, with a timing resolution of 161\,ps. The
normalized correlation function ($g^{(2)}(\tau)$)  shown in
Figure~\ref{fig:rabi} oscillates with a period that corresponds to the
Rabi frequency. Since $\Omega$\,$\propto$\,$\sqrt{P}$, for subsequent
measurements we determine $\Omega$  by measuring the the cooling beam
power $P$. 


\end{document}